\documentclass[12pt]{article}
\usepackage{graphicx,psfrag}

\newcommand{\bea}{\begin{eqnarray}}
\newcommand{\eea}{\end{eqnarray}}
\newcommand{\simgt}{\hbox{ \raise3pt\hbox to 0pt{$>$}\raise-3pt\hbox{$\sim$} }}
\newcommand{\simlt}{\hbox{ \raise3pt\hbox to 0pt{$<$}\raise-3pt\hbox{$\sim$} }}

\begin{document}
\begin{titlepage}
\title{\bf \Large
\vspace{28mm}
New Method for Reconstructing\\ Effective Top Quark Spin
\vspace{2cm}}
\author{Y.~Sumino$^a$ and S.~Tsuno$^b$
\\ \\ \\
\normalsize $^a$\,Department of Physics, Tohoku University,
Sendai, 980-8578 Japan\\
\normalsize 
$^b$\,Department of Physics, Okayama University, Okayama, 700-8530 Japan
}
\date{}
\maketitle
\thispagestyle{empty}
\vspace{-4.5truein}
\begin{flushright}
{\bf TU--761}\\
{\bf December 2005}
\end{flushright}
\vspace{4.5truein}
\begin{abstract}
\noindent
{\small
We propose a new method for reconstructing an effective spin direction
of a semi-leptonically decayed top quark.
The method is simple:  for instance, it does not require
the spin information of the antitop quark
in a $t\bar{t}$ event.
The reconstructed effective spin is expected to be useful 
in hadron collider experiments.
We demonstrate its usefulness in an analysis of
the top decay vertex.
}
\end{abstract}
\vfil

\end{titlepage}

After ten years since the discovery of the top quark 
\cite{Abe:1995hr}, 
as yet only limited experimental data on its properties are available,
and details of the nature of the top quark are still to be unveiled.
So far, there exists no evidence for significant deviations from the Standard Model (SM)
predictions concerning top quark properties.
On the other hand, since
the mass of the top quark approximates the electroweak symmetry-breaking
scale, there are hopes that symmetry-breaking physics may manifest itself
through non-standard interactions of the top quark.
For detailed examinations of top quark interactions, 
it is expected that the top quark spin can be used as a powerful
analysis tool.
This is because, in the SM (and in many of its extensions), 
the top quark decays before it
hadronizes, and the spin information of the top quark is directly reflected
to the distributions of its decay products.
Hence, we may utilize the top quark spin 
for disentangling different top quark interactions efficiently.
This is in contrast to the other lighter quarks, for which hadronization effects
dilute the spin information at the quark level severely.

In a future $e^+e^-$ linear collider experiment, we will be able
to utilize the spin of the top quark quite efficiently.
This is 
because produced top quarks will naturally be polarized, due to 
parity-violating nature of the interactions in the top quark production
process.
The top quark
polarization can even be controlled or raised to high values by 
tuning the polarization of the
initial electron beam.

By contrast, spin reconstruction of top quarks in hadron collider experiments
is a non-trivial task.
At Tevatron and LHC, top quarks are produced mainly through 
$t\bar{t}$ production processes, and these top quarks
are known to be hardly
polarized \cite{nlospin}.
Two types of methods have been proposed and studied for utilizing 
the top quark
spin at these colliders.
One is to take advantage of the correlation between the top quark spin and
antitop quark spin in the $t\bar{t}$ events.
The other method is to use polarized top quarks produced through
the single top production process.
Unfortunately, so far, not much information has been obtained
by applying these methods in analyzing
real top quark data in the Tevatron experiments,
due to intrinsic disadvantages of the methods.
The only analysis that has been performed
is a spin-correlation measurement
by D0, which put a fairly loose bound on a
correlation coefficient \cite{D0spincorrelation}.

Disadvantages in using the top-antitop spin correlation in the
$t\bar{t}$ events
are as follows.
If we analyze the $t\bar{t}$ events which decay in the $dilepton$ channel
(both $t$ and $\bar{t}$ decay semi-leptonically), data statistics is
low due to the small branching fraction.
Furthermore, there are two missing neutrino momenta in each event,
which make reconstruction of the event topology non-trivial.
Instead, if we analyze the $t\bar{t}$ events which decay in the $lepton+jets$ channel
(one of the top quarks decays semi-leptonically and the other decays hadronically),
we have more statistics, but reconstruction of the spin of the hadronically-decayed
top quark needs to go through complicated procedures.
The reconstruction process is affected significantly by kinematical cuts, 
and often important information
is lost by the cuts.
Accurate estimation of the effects of kinematical cuts and
event reconstruction is crucial as well.
The complex procedures bring in sizable systematic uncertainties
in the top spin reconstruction.

Disadvantages in using the single top production
process are that there are huge background cross sections for 
$Wb\bar{b}$ and $Wb\bar{b}+jets$ processes, and that 
these background cross sections have not been estimated accurately.
In fact, up to now the single top production process
has not been observed at Tevatron,  as opposed to original expectations, 
due to lack of data statistics and difficulty
in the background estimation.

We would expect that, when a huge top quark sample is available at LHC,
these difficulties will eventually be overcome, along with reduction of
statistical errors as well as better understandings of systematic
uncertainties; see e.g.\ \cite{Atwood:1992vj}.
On the other hand, it is certainly desirable to develop another
method for top spin reconstruction that can be applied to
a high statistics sample, and that involves small and controlled systematic
uncertainties.
In this paper, we propose a new method that can meet such 
criteria.
This method can be applied to a semi-leptonically decayed top quark,
without requiring reconstruction of the spin of the antitop quark.
We will demonstrate usefulness of our method in an analysis of the top-quark
decay vertex.
A more detailed application of our method is given in \cite{fullpaper}, where
sensitivities to anomalous couplings in the top decay vertex
are studied, using our method, and
taking into account realistic experimental conditions expected at 
Tevatron and LHC.
There, it is shown that our method for effective top spin reconstruction
is indeed practically useful.

One may be perplexed, since there is no spin vector (polarization vector) 
associated with
an unpolarized state; 
one may well argue that it is impossible to reconstruct the spin of an 
unpolarized top quark.
While this argument is correct on its own,
we can still reconstruct an effective ``spin direction'' of a top quark, which is
practically useful in analyses of top decays.

The method we propose is simple and naive.
It is based on the following two well-known facts:
\begin{enumerate}
\renewcommand{\labelenumi}{(\roman{enumi})}
\item
Within the SM, the charged lepton in the semi-leptonic
decay of a 100\% polarized top quark 
is emitted preferentially in the top quark spin direction.
In fact, at tree level of the SM, 
the normalized angular distribution of the lepton is given by \cite{langdist1}
\bea
\frac{1}{\cal N} \, 
\frac{d\Gamma(t \to b l \nu)}{d \cos \Theta}
= \frac{1+\cos\Theta}{2} ,
\label{leptonangdist}
\eea
where $\Theta$ denotes the angle between the top spin direction and the
lepton direction in the top rest frame.
$\cal N$ represents the normalization constant such that the integral
upon $\cos\Theta$ becomes unity.
It is known that the one-loop QCD correction hardly modifies
the above angular distribution 
\cite{langdist2}.
\item
If we include anomalous couplings in the top decay vertex,
their effects on the lepton
angular distribution enter only from quadratic dependences
\cite{hioki}.
(All terms linear in the anomalous couplings vanish.)
Namely, when the anomalous couplings are small, their effects are very
suppressed.
\end{enumerate}
Unpolarized top quarks can be interpreted as an admixture, where one-half of
them have their spins in $+\vec{n}$ direction and the other half have
their spins in $-\vec{n}$ direction, for an arbitrary
chosen unit vector $\vec{n}$.
The directions of the charged leptons from the top quarks
with $\pm \vec{n}$ spin are emitted preferentially in the 
$\pm \vec{n}$ direction in the top rest frame.
Hence, it seems reasonable to project the 
direction of the lepton $\vec{p}_l/|\vec{p}_l|$ onto the $\vec{n}$-axis
and define an effective spin direction as 
${\rm sign} (\vec{n}\!\cdot\!\vec{p}_l) \times \vec{n}$
for each event.
According to eq.~(\ref{leptonangdist}), in this way we choose
the correct direction with probability 75\% on average.
That we can choose any axis $\vec{n}$, and that any choice
is equivalent (if we ignore experimental environment), guarantee the rotational
invariance of the unpolarized state of the top quark.
Due to the above property (ii), the defined direction is hardly affected
by the anomalous couplings in the top decay vertex if they are small,
so that it is appropriate for an effective spin direction.

Importance of this definition consists in our finding that certain
angular distributions of the top 
decay products 
with respect to the effective spin direction
reproduce fairly well the corresponding angular distributions from a
truly polarized top quark.
This is the case even including anomalous couplings in the
top decay vertex.
This is the main aspect to be addressed in this article.

Provided that produced top quarks are perfectly unpolarized,
and provided
that we disregard effects by
kinematical cuts and acceptance corrections, 
there is no difference on which
spin axis $\vec{n}$ we choose to project the direction of the charged lepton.
In most part of the paper, we consider this ideal case.
We will briefly discuss effects of incorporating realistic experimental conditions
at the end.

We start by explaining our setup of the top decay vertex including form factors.
We assume that deviations of the top decay form factors from the
tree-level SM values are small.
Then we consider only those form factors which induce
deviations of the differential distributions of top decay products
at the first order in the anomalous
form factors.
That leaves only two form factors
 $f_1^L$ and $f_2^R$ in the limit
$m_b \to 0$ and for onshell $W$,
although the most general $tbW$ vertex includes
six independent form factors \cite{cpyuan}:
\bea
\Gamma^\mu_{Wtb} = -\frac{g_W}{\sqrt{2}} \, V_{tb} \, 
\bar{u}(p_b) \left[
\gamma^\mu \, f_1^L \, P_L -
\frac{i \, \sigma^{\mu\nu}k_\nu}{M_W} \, f_2^R \, P_R 
\right]
u(p_t) ,
\eea
where $P_{L,R}=(1 \mp \gamma_5)/2$ are the 
left-handed/right-handed projection operators;
$k$ is the momentum of $W$.
For simplicity, we further assume that $f_1^L$ and $f_2^R$
are real.\footnote{
We note that the anomalous couplings for the right-handed bottom quark
and the $CP$-odd anomalous couplings,
which we neglect here, are severely constrained indirectly from the
measurements of $Zb\bar{b}$ vertex at LEP and of $b \to s\gamma$ process
\cite{const2}.
This may provide another justification for neglecting
these form factors in our simplified analysis.
} 
At tree level of the SM, $f_1^L=1$ and $f_2^R=0$.
We will be concerned only with the top decay process $t \to bW$,
where the $Q^2$ value is fixed, therefore, we treat the form factors
as constants (couplings) henceforth.

\begin{figure}[t]
\begin{center}
\psfrag{AW}{$W$}
\psfrag{Atop}{top}
\psfrag{Aspin}{spin}
\psfrag{WRestFrame}{$W$ rest frame}
\psfrag{ThetaW}{$\theta_{W}$}
\psfrag{ThetaL}{$\theta_{l}$}
\psfrag{PhiL}{$\phi_{l}$}
\psfrag{eL}{$l$}
\includegraphics[width=4.0cm]{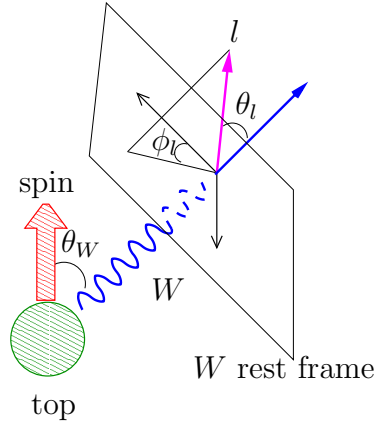} 
\caption{\small Schematic view of the angle definitions.
}
\label{angle_def}
\end{center}
\end{figure}

Using the above decay vertex and taking the narrow
width limit for $W$, the differential decay 
distribution of $W$ and $l$ in the semi-leptonic decay of a 
100\% polarized top quark is given by \cite{cpyuan}
\begin{eqnarray}
\frac{d\Gamma(t \rightarrow bW \! \rightarrow  bl\nu)}
{d\cos\theta_{W}d\cos\theta_{l}d\phi_{l}} =
A 
{\left|
(f_2^R+f_1^L\frac{m_t}{M_{W}})\cos\frac{\theta_{W}}{2}\sin\theta_{l}
+2\, (f_1^L+f_2^R\frac{m_t}{M_{W}})e^{-i\phi_{l}}\sin \frac{\theta_{W}}{2}
\sin^2\frac{\theta_{l} }{2}
\right|}^{2} ,
\nonumber\\
\label{tripleangdist-truespin}
\end{eqnarray}
with
\begin{equation}
A = \frac{3G_{F}|V_{tb}|^{2}M_{W}^{2}(m_t^{2}-M_{W}^{2})^{2}}
{32\sqrt{2}\,\pi\, m_t^{3}} \times Br(W \rightarrow l\nu) \quad .
\label{eq6}
\end{equation}
Here, $G_{F}$ is the Fermi constant.
$\theta_{W}$ is defined as the angle between the top polarization vector 
and the direction of $W$ in the top quark rest frame. 
$\theta_{l}$ is defined as the lepton 
helicity angle, which is the angle of the charged lepton in the rest frame of 
$W$ with respect to the original direction of the travel of $W$. 
$\phi_{l}$ is 
defined as the azimuthal angle of $l$ around the original direction of 
the travel of $W$.
A schematic view of the angle definitions is shown in 
Fig.~\ref{angle_def}. 
The above differential distribution contains fully differential information
on the decay $t \rightarrow bW \! \rightarrow  bl\nu$.

The corresponding differential distribution with respect to the
effective spin direction 
can be computed in the following way.
An arbitrary unit vector $\vec{n}$ is chosen as the spin axis
in the top rest frame.
We denote the charged lepton momentum as $p_l$ in the same frame.
Then, if ${\vec{n}\cdot\vec{p}_l}>0$, we define the
effective spin vector 
to be $\vec{s}_{\rm eff}=\vec{n}$, whereas if ${\vec{n}\cdot\vec{p}_l}<0$,
we define the
effective spin vector 
 to be $\vec{s}_{\rm eff}=-\vec{n}$.
The angle $\Theta_{\rm eff}$ between $\vec{s}_{\rm eff}$ and $\vec{p}_l$
is given by
\bea
&&
\cos \Theta_{\rm eff} \equiv
\frac{\vec{s}_{\rm eff}\cdot\vec{p}_l}{|\vec{p}_l|}=
\frac{\sqrt{1-\beta_W^2}}{1+\beta_W \cos\theta_l}
\left(
\sin \theta_l \cos \phi_l \sin \theta_W
+ \frac{\cos \theta_l + \beta_W}{\sqrt{1-\beta_W^2}} \cos\theta_W
\right) ,
~~~~~~~
\eea
where $\beta_W=({m_t^2-M_W^2})/({m_t^2+M_W^2}) $
denotes the velocity of $W$ in the top rest frame.
$\theta_W$ and $\phi_l$ are defined as in Fig.~\ref{angle_def}
with respect to $\vec{s}_{\rm eff}$ (instead of the top polarization vector).\footnote{
It would be more accurate to denote these angles as
$\theta_{W,{\rm eff}}$ and $\phi_{l,{\rm eff}}$, but to avoid illegibility
we use the same notation as in eq.~(\ref{tripleangdist-truespin}).
}
Thus,
\bea
\left[
\frac{d\Gamma(t \rightarrow bW \! \rightarrow  bl\nu)}
{d\cos\theta_{W}d\cos\theta_{l}d\phi_{l}} 
\right]_{\rm eff.\, spin} &=&
\left[
\frac{d\Gamma(t \rightarrow bW \! \rightarrow  bl\nu)}
{d\cos\theta_{W}d\cos\theta_{l}d\phi_{l}} 
\right]_{\rm unpol.} \times
2\, \theta(\cos \Theta_{\rm eff}) .
\label{tripleangdist-effspin}
\eea
$\theta(x)$ represents the unit step function.
Here,
the decay distribution from an {\it unpolarized} top quark is given by
\bea
\left[
\frac{d\Gamma(t \rightarrow bW \! \rightarrow  bl\nu)}
{d\cos\theta_{W}d\cos\theta_{l}d\phi_{l}} 
\right]_{\rm unpol.} = 
\frac{1}{4}\,A \,
\biggl[
\Bigl(f_1^L\frac{m_t}{M_W}+f_2^R \Bigr)^2
\, \sin^2 \theta_l
+ 4 \, \Bigl(f_1^L +f_2^R\frac{m_t}{M_W} \Bigr)^2
\,\sin^4  \frac{\theta_l}{2} 
\biggr] \, .
\eea
Obviously, it is independent of $\theta_W$ and $\phi_l$, since there is
no reference spin vector.
Therefore, the dependences on $\theta_W$ and $\phi_l$ of the
differential distribution with respect to the effective spin direction enter
only through the step function 
$\theta(\cos \Theta_{\rm eff})$
on the right-hand-side of eq.~(\ref{tripleangdist-effspin}).

At this fully differential level, 
$[d\Gamma/d\cos\theta_Wd\cos\theta_ld\phi_l]_{\rm eff.\, spin}$
[eq.~(\ref{tripleangdist-effspin})]
is only a crude approximation to
$d\Gamma/d\cos\theta_Wd\cos\theta_ld\phi_l$
[eq.~(\ref{tripleangdist-truespin})].
It can be seen, for instance, from the existence of the step function or
from the factorized form of the dependences on $(f_1^L,f_2^R)$ and on
$(\theta_W,\phi_l)$ in eq.~(\ref{tripleangdist-effspin}), neither of which
is in the structure of eq.~(\ref{tripleangdist-truespin}).

\begin{figure}[t]
\begin{center}
\begin{tabular}{ccc}
\psfrag{cosl}{$\cos\theta_l$}
\psfrag{cosw}{$\cos\theta_W$}
\includegraphics[width=7.0cm]{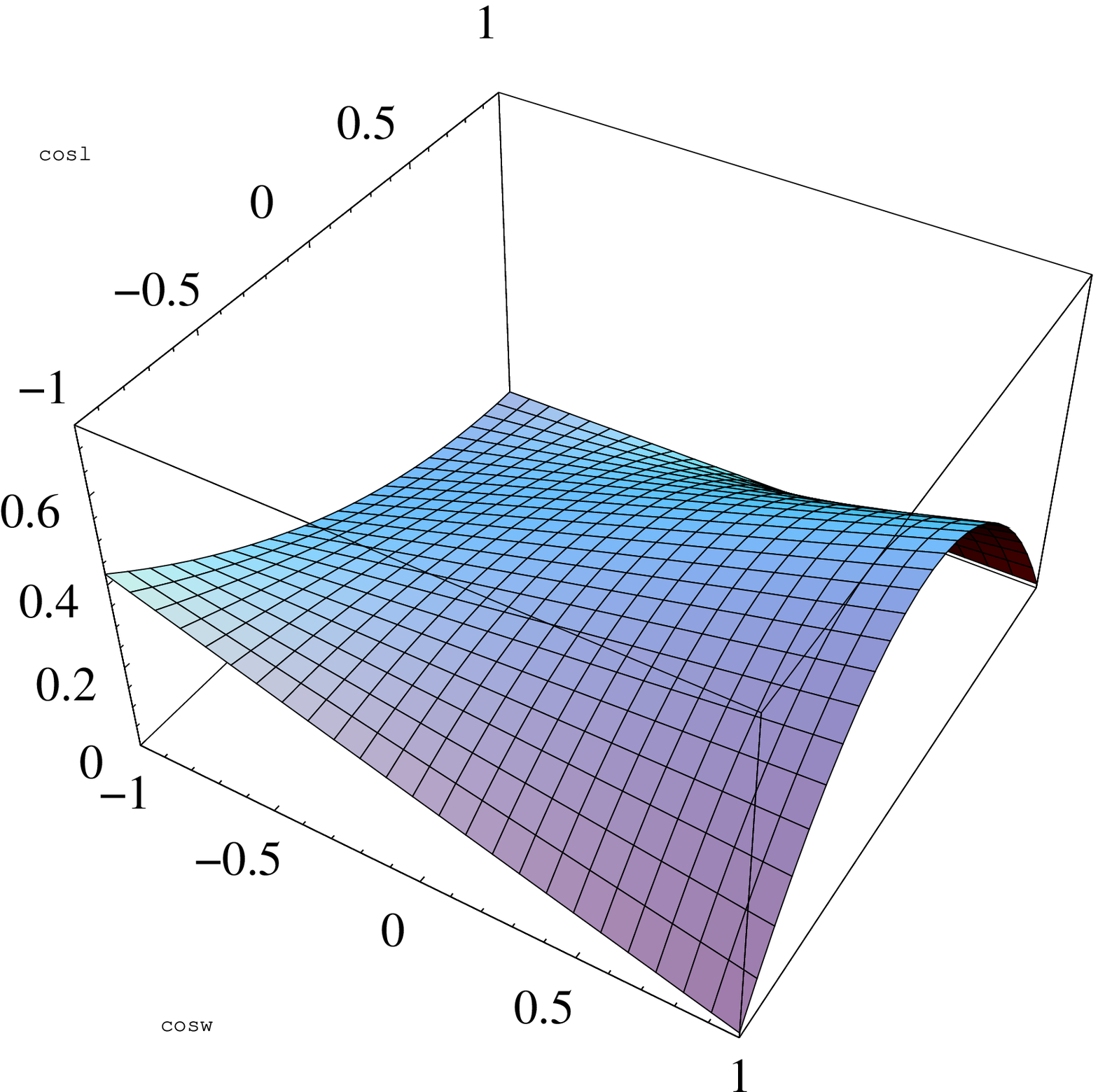} &&
\psfrag{cosl}{$\cos\theta_l$}
\psfrag{cosw}{$\cos\theta_W$}
\includegraphics[width=7.0cm]{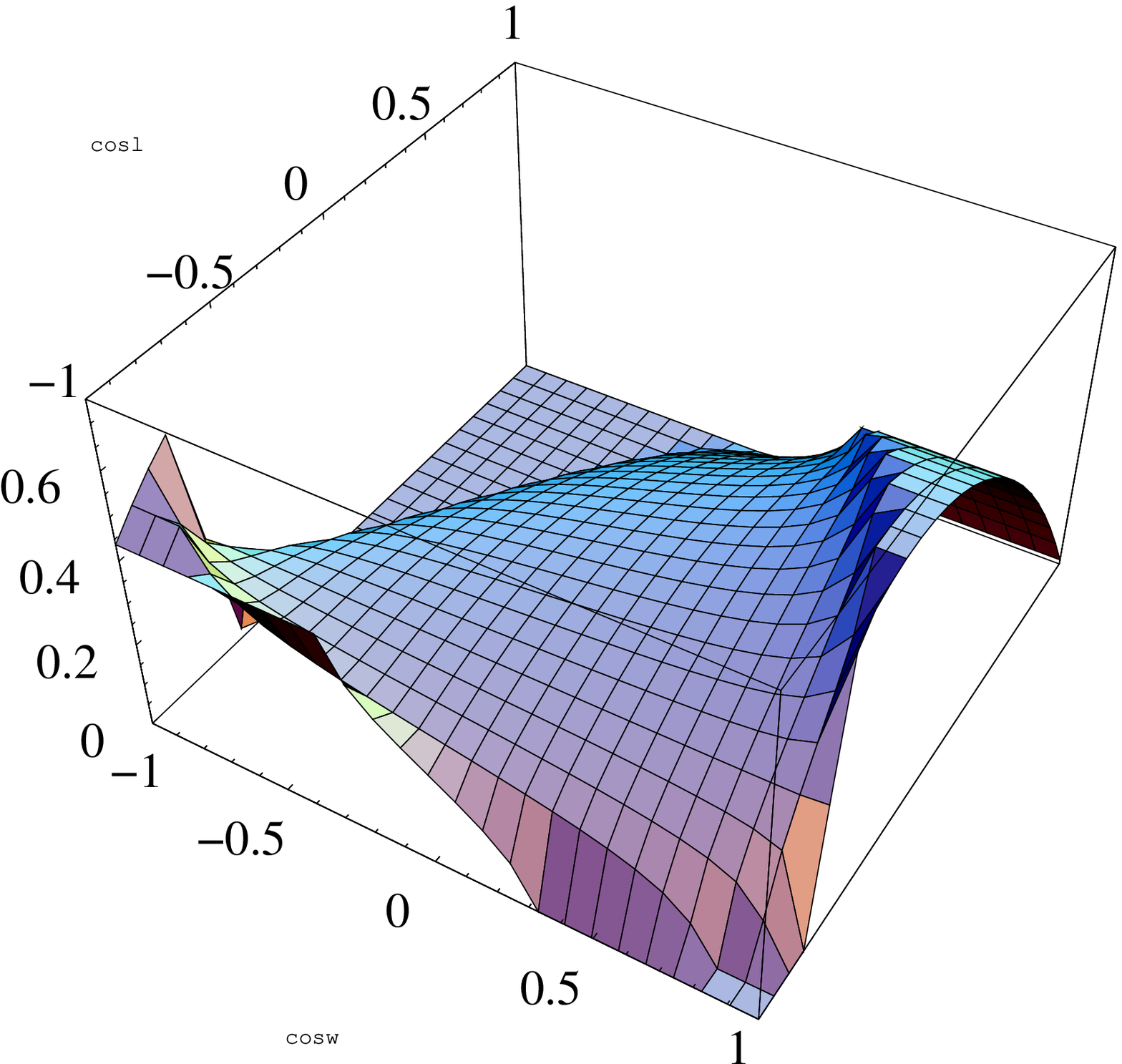}\\
(a) && (b)
\end{tabular}
\caption{\small Normalized double
angular distributions for $(f_1,f_2) =(1,0)$
(a) using the true spin direction 
and (b) using the effective spin direction,
corresponding to eqs.~(\ref{doubleangdistr-truespin}) and
(\ref{doubleangdistr-effspin}), respectively.
They are normalized to unity upon integration.}
\label{double-ang-dist}
\end{center}
\end{figure}

Let us integrate over $\phi_l$ and compare the double angular distributions
with respect to the true and effective spin directions:\\
\bea
&&
\frac{
d\Gamma(t \rightarrow bW \! \rightarrow  bl\nu)
}{
d\cos\theta_{W}d\cos\theta_{l}
} =
\pi \, A \,
\biggl[
\Bigl(f_1^L\frac{m_t}{M_W}+f_2^R \Bigr)^2
\cos^2 \frac{\theta_W}{2} \, \sin^2 \theta_l
\nonumber \\
&& ~~~~~~~~~~~~~~~~~~~~~~~~~~~~~~~~~~~~~~~~~
+ 4 \, \Bigl(f_1^L +f_2^R\frac{m_t}{M_W} \Bigr)^2
\, \sin^2 \frac{\theta_W}{2} \,\sin^4  \frac{\theta_l}{2} 
\biggr] \, ,
\label{doubleangdistr-truespin}
\\ &&
\left[
\frac{d\Gamma(t \rightarrow bW \! \rightarrow  bl\nu)}
{d\cos\theta_{W}d\cos\theta_{l}} 
\right]_{\rm eff.\, spin} 
=
\left[
\frac{d\Gamma(t \rightarrow bW \! \rightarrow  bl\nu)}
{d\cos\theta_{W}d\cos\theta_{l}d\phi_{l}} 
\right]_{\rm unpol.} \times
2\, g(y) \, ,
\label{doubleangdistr-effspin}
\eea
where
\bea
&&
y = -\frac{\cos \theta_l + \beta_W}{\sqrt{1-\beta_W^2}}\,
\frac{\cot \theta_W}{\sin \theta_l} ,
~~~~~~~~~~~~
g(x) = \left\{
\begin{array}{ll}
0&\mbox{if $x \geq 1$}\\
2\pi & \mbox{if $x \leq -1$}\\
\pi - 2 \arcsin x & \mbox{if $-1<x<1$}
\end{array}
\right. .
\eea
Numerically these two distributions become reasonably close to each 
other.\footnote{
It is not obvious from the explicit formulas for the distributions.
In particular, eq.~(\ref{doubleangdistr-effspin})
still has a factorized form concerning
the dependences on $(f_1^L,f_2^R)$ and 
$\theta_W$, which is different from eq.~(\ref{doubleangdistr-truespin}).
}
This is demonstrated in Figs.~\ref{double-ang-dist}(a)(b), 
in which both double angular distributions
are displayed for $(f_1^L,f_2^R)=(1,0)$ (tree-level SM).
The distributions are normalized to unity upon integration.
Qualitative features of the
bulk distribution shape of
$d\Gamma/d\cos\theta_Wd\cos\theta_l$ are
reproduced by $[d\Gamma/d\cos\theta_Wd\cos\theta_l]_{\rm eff.\, spin}$.

It has been known that 
the double angular distribution $d\Gamma/d\cos\theta_Wd\cos\theta_l$
is useful for probing the anomalous coupling 
$f_2^R$ \cite{ikematsu}.
To see sensitivities to the anomalous couplings semi-quantitatively, 
we divide the phase space into four regions as
\begin{equation}
\begin{array}{lrccrcl}
\mathrm{Region}\ A : \quad & -1 \; & \leq \; \cos\theta_{W} \; \leq \; 0 
& \quad \mathrm{and} \quad & -1 \; & \leq \; \cos\theta_{l} \; \leq \; 0 & , \\
\mathrm{Region}\ B  : \quad & -1 \; & \leq \; \cos\theta_{W} \; \leq \; 0
& \quad \mathrm{and} \quad & 0 \; & \leq \; \cos\theta_{l} \; \leq \; 1 &, \\
\mathrm{Region}\ C  : \quad & 0 \; & \leq \; \cos\theta_{W} \; \leq \; 1 
& \quad \mathrm{and} \quad & -1 \; & \leq \; \cos\theta_{l} \; \leq \; 0 & ,\\
\mathrm{Region}\ D  : \quad & 0 \; & \leq \; \cos\theta_{W} \; \leq \; 1 
& \quad \mathrm{and} \quad & 0 \; & \leq \; \cos\theta_{l} \; \leq \; 1 &  ,\\
\end{array}
\label{eq11}
\end{equation}
and define the event fraction in each region by
\bea
&&
R_i = {\cal N}^{-1} \int_{{\rm Region}\, i} d\cos\theta_Wd\cos\theta_l \,
\frac{d\Gamma(t \to bW \to bl\nu)}{d\cos\theta_Wd\cos\theta_l} ,
\eea
where
\bea
&&
{\cal N}=\int_{-1}^1 d\cos\theta_W \int_{-1}^1 d\cos\theta_l \,
\frac{d\Gamma(t \to bW \to bl\nu)}{d\cos\theta_Wd\cos\theta_l} 
\eea
represents the top-quark partial width to $bl\nu$.
Each $R_i$ is a function of $f_2^R/f_1^L$.
We also define the event fractions
$R_i^{\rm eff}$ in the same manner
using the effective spin direction instead of the
true spin direction.

\begin{figure}[t]
\begin{center}
\begin{tabular}{ccc}
\psfrag{RA}{$R_A$}
\psfrag{RB}{$R_B$}
\psfrag{RC}{$R_C$}
\psfrag{RD}{$R_D$}
\psfrag{ratio}{\footnotesize $f_2^R/f_1^L$}
\includegraphics[width=7.5cm]{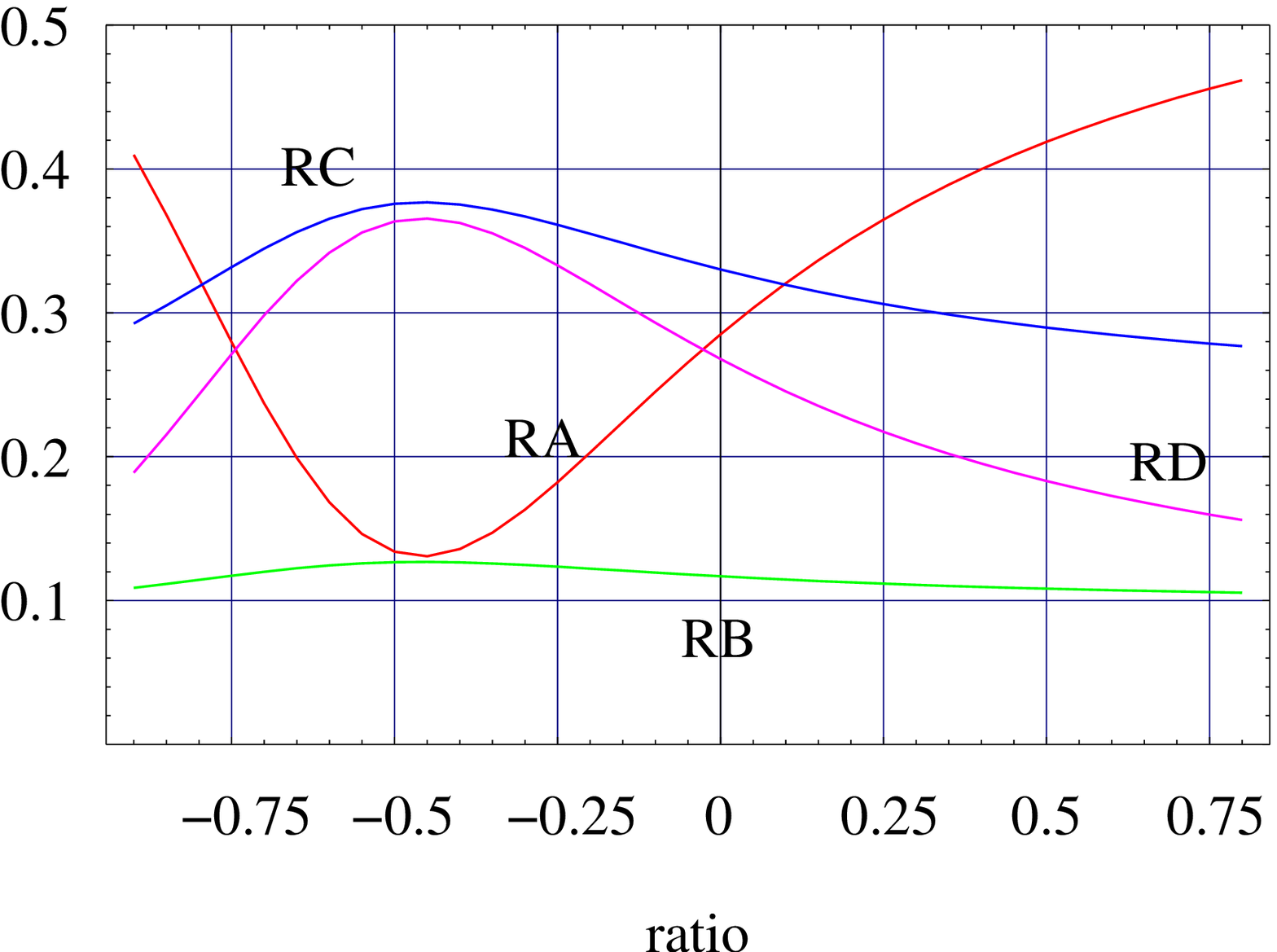} &~~~~~&
\psfrag{RA}{$R_A^{\rm eff}$}
\psfrag{RB}{$R_B^{\rm eff}$}
\psfrag{RC}{$R_C^{\rm eff}$}
\psfrag{RD}{$R_D^{\rm eff}$}
\psfrag{ratio}{\footnotesize $f_2^R/f_1^L$}
\includegraphics[width=7.5cm]{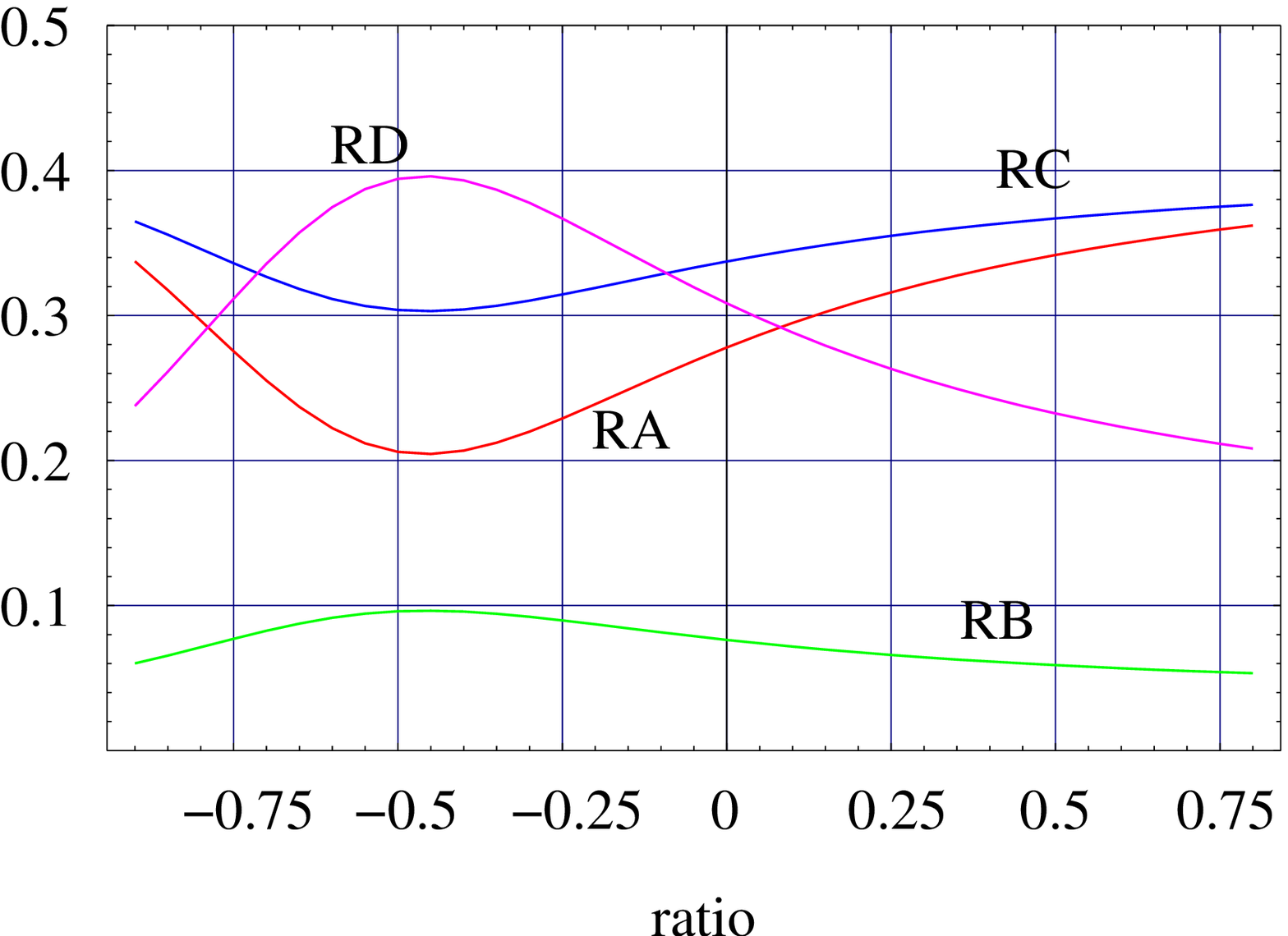}
\vspace{3mm}
\\
~~~~~(a) &~~~& ~~~~~(b)
\end{tabular}
\caption{\small 
Event fractions in the four regions  as functions of $f_2^R/f_1^L$
(a) using the true spin direction ($R_i$)
and (b) using the effective spin direction ($R_i^{\rm eff}$).
}
\label{Ri}
\end{center}
\end{figure}

We compare the dependences of $R_i$ and $R_i^{\rm eff}$
on $f_2^R/f_1^L$ in Figs.~\ref{Ri}(a)(b).
From the figures, we see that major features of the 
$f_2^R/f_1^L$ dependences of 
$R_i$ are reproduced by $R_i^{\rm eff}$.
In fact, the dependences of $R_i^{\rm eff}$ are consistent with the
observation that, if we use the effective spin direction, we misidentify
the correct spin direction with 25\% probability on average.
Namely, we misidentify
Region $A$ with $C$, and Region $B$ with $D$,
so if we combine $R_i$s in Fig.~\ref{Ri}(a) reweighting them
with this misidentification probability, we
obtain the curves similar to those plotted in Fig.~\ref{Ri}(b).
Since the $f_2^R/f_1^L$ dependence of the most sensitive event
fraction
$R_A^{\rm eff}$ is
about half of that of $R_A$, if we use the effective spin direction,
we would expect a sensitivity to
$f_2^R/f_1^L$ roughly half of what would be obtained with the true spin direction.
A closer examination of the sensitivities to $f_2^R/f_1^L$ incorporating realistic
experimental conditions is given in \cite{fullpaper}.

The one-loop QCD correction to $d\Gamma/d\cos\theta_Wd\cos\theta_l$
has been computed in \cite{QCD1Loop}.
A large part of the correction goes to a variation of the normalization
of the partial decay width, which amounts to about 9\%.
On the other hand, the correction to the
normalized double angular distribution
is at the level of 1--2\% or less.
Although the one-loop QCD correction to 
$[d\Gamma/d\cos\theta_Wd\cos\theta_l]_{\rm eff.\, spin}$
has not been computed yet, we expect that
it would not be
very different from the correction to $d\Gamma/d\cos\theta_Wd\cos\theta_l$.
If this is so,
we may be able to measure the QCD correction 
to the normalized double angular distribution
at LHC, provided that a good understanding of systematic uncertainties is possible;
cf.\ \cite{fullpaper}.

\begin{figure}[t]
\begin{center}
\psfrag{W}{$W$}
\psfrag{b}{$b$}
\psfrag{nu}{$\nu$}
\psfrag{cosi}{$\cos\theta_i$}
\includegraphics[width=7.0cm]{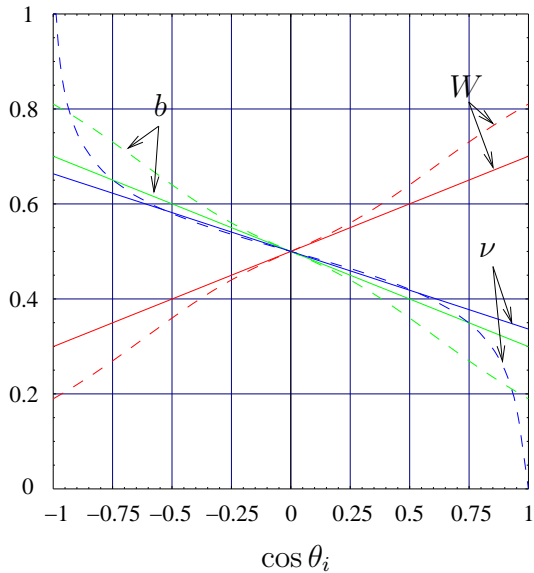} 
\caption{\small 
Normalized angular distributions for $b$,$W$,$\nu$
in the top rest frame for $(f_1^L,f_2^R)=(1,0)$,
using the true spin direction (solid lines) and using the
effective spin directions (dashed lines).
}
\label{angular-distr}
\end{center}
\end{figure}

We may also compare the angular distributions
$d\Gamma/d\cos\theta_i$ and $[d\Gamma/d\cos\theta_i]_{\rm eff.\, spin}$, 
where $\theta_i$ denotes the
angle between the direction of
particle $i$ and the top polarization vector or the effective
spin direction $\vec{s}_{\rm eff}$ in the top rest frame.
The normalized angular
distributions are shown in Fig.~\ref{angular-distr} 
for $i=b,W,\nu$ and $(f_1^L,f_2^R)=(1,0)$.
(The lepton angular distributions are trivial, 
so we do not show them here.)
The angular distributions with respect to the true spin direction
depend linearly on $\cos\theta_i$.
In this case, it is customary to parametrize a normalized
angular distribution by 
$\frac{1}{2}(1+\alpha_i\cos\theta_i)$ and refer to
$\alpha_i$ as a correlation coefficient.
Since $b$ and $W$ are emitted back-to-back in the top rest frame,
$\alpha_b=-\alpha_W$.
The correlation coefficients $\alpha_i$ 
corresponding to the true spin direction have been computed in \cite{angularcorr}.
On the other hand, the angular distributions with respect to $\vec{s}_{\rm eff}$ 
are not given by linear functions of $\cos\theta_i$.
Their analytic expressions are complicated, which we do not present here.
Numerically, the angular distributions for $b$ and $W$ are close to
linear shape, 
while that of $\nu$ is considerably different from linear shape close
to $\cos\theta_\nu=\pm1$.

If we approximate $[d\Gamma/d\cos\theta_i]_{\rm eff.\, spin}$
for $i=b,W$ by linear functions of $\cos\theta_i$,
the correlation coefficients
$|\alpha_b|$ and $|\alpha_W|$ for the effective spin direction are about twice larger than 
those for the true top spin direction.
As for the angular distribution of $\nu$ with respect to
$\vec{s}_{\rm eff}$, on average the
slope of the distribution
is steeper (the correlation between the neutrino direction and 
$\vec{s}_{\rm eff}$ is stronger)
than that of $d\Gamma/d\cos\theta_\nu$.
These enhancements in the angular
correlations, if we use the effective spin direction
instead of the true spin direction,
stem from purely kinematical origins.
It can be understood as follows.
Consider a hypothetical case, in which no correlation between
the true spin direction and direction of $W$ exists (the decay is isotropic).
Even in this case, there is a positive correlation between
the effective spin direction and $W$ in 
the top rest frame, since the charged
lepton is emitted more in the direction of $W$ 
due to the boost by $W$.
Similarly, (hypothetically)
even in the absence of any correlation between
the true spin direction and neutrino direction,
there is a negative correlation between the
lepton direction and neutrino direction in the top rest frame,
since they are 100\% anticorrelated (back-to-back) in the $W$ rest frame.
These kinematical effects bias the angular correlations 
to be stronger if we use the effective spin direction.

We also examined the $f_2^R/f_1^L$ dependences of the
angular distributions 
$d\Gamma/d\cos\theta_i$ and $[d\Gamma/d\cos\theta_i]_{\rm eff.\, spin}$.
The $f_2^R/f_1^L$ dependences of 
the latter distributions are much weaker than 
those of the former distributions.
The $f_2^R/f_1^L$ dependences of $[d\Gamma/d\cos\theta_i]_{\rm eff.\, spin}$
for $i=b,W$
are reduced as compared to the $f_2^R/f_1^L$ dependences of 
the double angular distribution
eq.~(\ref{doubleangdistr-effspin}).
This is
due to cancellations of $f_2^R/f_1^L$ dependences between
$R_A^{\rm eff}$ and $R_B^{\rm eff}$, and between 
$R_C^{\rm eff}$ and $R_D^{\rm eff}$; see Fig.~\ref{Ri}(b).
Insensitivity of $[d\Gamma/d\cos\theta_\nu]_{\rm eff.\, spin}$ 
to $f_2^R/f_1^L$ stems
from a strong (anti)correlation between the effective spin direction 
(or  the lepton direction) and
the $\nu$ direction.
Since the $f_2^R/f_1^L$ dependences
of $[d\Gamma/d\cos\theta_i]_{\rm eff.\, spin}$ are weak, it is much
more advantageous to use the double angular distributions
[eq.~(\ref{doubleangdistr-effspin})]
or event fractions $R_i^{\rm eff}$ for gaining sensitivities to
$f_2^R/f_1^L$.

Up to now, in defining the effective spin direction,
we assumed that the initial top quark is completely 
unpolarized and neglected experimental environment.
In practice, under realistic experimental conditions,
different choices of spin basis (axis) $\vec{n}$ lead to different 
distributions of decay products.
Effects of kinematical cuts are by far the largest.
Based on detailed Monte Carlo simulation studies incorporating
realistic experimental conditions expected at Tevatron and LHC,
it is found that the top helicity axis
$\vec{p}_t/|\vec{p}_t|$ defined in the $t\bar{t}$ c.m.\ frame
(opposite of the direction of $\bar{t}$ in the top rest frame)
is an optimal choice for the spin axis $\vec{n}$.
Other choices, such as beamline axis and 
the off-diagonal spin basis \cite{offdiagonal}, turn out to be inappropriate,
since the
original distributions [e.g.\ Fig.~\ref{double-ang-dist}(b)] 
are strongly distorted
by the effects of kinematical cuts,
and also because the sensitivities to the anomalous
couplings are much reduced.
This can be understood as follows.
If we choose the beamline axis,
the small $E_{T}$ (transverse energy) and large $|\eta|$ 
(pseudorapidity) regions correspond
to the regions $\cos \theta_W \simeq \pm 1$,
and events that fall into these kinematical regions are rejected
by cuts such as the requirements for the minimum 
transeverse energy ($E_T$ cut) or acceptance correction ($|\eta|$ cut)
for the lepton and jets.
In particular, events in the kinematical 
regions most sensitive to a variation of $f_2^R/f_1^L$,
close to 
$(\cos \theta_W,\cos \theta_l )= (1,1)$ and $(-1,-1)$,
are lost.
At Tevatron, the status of the off-diagonal basis is similar 
to the beamline axis, since the off-diagonal basis is not very different
from the beamline axis.
(At LHC, there is no good definition of the off-diagonal basis.)
On the other hand, 
if we choose the top helicity axis, after integrating over 
all top quark directions, effects of the cuts are averaged over and
no significant distortion from the original distribution is found.
See \cite{fullpaper} for details.

As is clear from the above definition,
our method can be applied not only to the hadron collider experiments
but also to a future $e^+e^-$ collider experiment.
Nevertheless, the primary motivation of our proposal is to use this method
at the current Tevatron experiment and at LHC.

In summary, we proposed to reconstruct an effective 
spin direction of a semi-leptonically decayed top quark
as the projection of the lepton direction onto an arbitrary
chosen axis in the top rest frame.
The reconstruction method is simple so that it would be feasible
in hadron collider experiments.
We demonstrated that this spin direction can be used to
probe anomalous couplings in the top decay vertex,
through measurements of a double angular distribution
or event fractions $R_i^{\rm eff}$.
Under realistic experimental conditions,
the top helicity axis seems to be an optimal choice for the spin axis.

\end{document}